\documentstyle[prb,aps]{revtex}
\newcommand{\eqb}{\begin{eqnarray}}
\newcommand{\eqe}{\end{eqnarray}}
\newcommand{\md}{\mbox{d}}
\newcommand{\nn}{\nonumber}
\begin{document}
\draft
\thispagestyle{headings}
\title{Comment on: Diffusion through a slab}
\author{U.~D.~J.~Gieseler \& J.~G.~Kirk}
\address{Max-Planck-Institut f\"ur Kernphysik, 
 Postfach 10 39 80, D-69029 Heidelberg, Germany }
\date{Received 2 December 1996; accepted for publication 20 January 1997}
\maketitle
\begin{abstract}
Mahan [J.~Math.~Phys.~{\bf 36}, 6758 (1995)] has calculated the transmission 
coefficient and angular distribution of particles which enter a {\em thick} 
slab at {\em normal} 
incidence and which diffuse in the slab with linear anisotropic, 
non-absorbing, scattering. Using orthogonality relations derived by McCormick
\& Ku\v{s}\v{c}er [J.~Math.~Phys.~{\bf 6}, 1939 (1965); {\bf 7}, 2036 (1966)]
for the eigenfunctions of 
the problem, this calculation is generalised to a 
boundary condition with particle input at {\em arbitrary}
angles. It is also shown how to use the orthogonality relations
to relax in a simple way the restriction to a thick slab.
\end{abstract}
We consider the equation of radiative transfer with anisotropic
scattering in a uniform slab, which occupies the space $0<z<D$,
together with a boundary condition 
which allows particles to enter the slab through the surface
$z=0$ at an angle $\theta = \arccos\mu_0$ to the normal:\cite{M46}
\eqb\label{transport}
\mu{\partial \over\partial z}f(z,\mu) + f(z,\mu) &=&
        {1\over2}\int\limits_{-1}^1\,{\rm d}\mu' f(z,\mu')
     +{3\over2}\,\mu\,g_1\int\limits_{-1}^1\,{\rm d}\mu' \mu'f(z,\mu')\,,
\eqe
\eqb\label{boundary}
\left.
\begin{array}{lcl}
   f(0,\mu) & = & 2\,\delta (\mu - \mu_0) \\
   f(D,-\mu) & = & 0                       
\end{array}
\right\} \,\, \mbox{for} \,\,\, \mu >0 \,.
\eqe
For thick slabs ($D\gg1$), 
Mahan \cite{Mahan} has presented a 
solution to this problem which is valid only for
$\mu_0=1$. Generalisation to 
arbitrary $\mu_0$ is of interest when, for example, the particles
which enter the slab come from a point source at finite distance, 
or diffuse before entering the slab. 
These problems require an
integration over the range of incident angles. 
Even for collimated beams, the experimental situation is generally 
one in which the particles are not normally incident. 
Mahan's method
is not readily generalised to solve this problem: 
his Eq.~(78) does
not hold when  $\mu_0\ne 1$, since then $A^{-1}(\mu_0)\neq 0$.

The general solution to Eq.~(\ref{transport}) is\cite{M8889}
\eqb\label{general}
  f(z,\mu) &=& a_s + 3 j\,\bigr[\mu - z(1-g_1)\bigl]
+\int\limits_0^1\md \nu
                \left\{\frac{M_L(\nu)}{\nu-\mu}e^{-z/\nu} 
                       + \delta(\nu-\mu)A(\nu)M_L(\nu)e^{-z/\nu}\right\} \nn \\ 
           & & +\int\limits_{-1}^0\md \nu
                \left\{\frac{M_R(\nu)}{\nu-\mu}e^{(D-z)/\nu} 
                       + \delta(\nu-\mu)A(\nu)M_R(\nu)e^{(D-z)/\nu}\right\}\,,
\eqe
where the constants $a_s$ and $j$, and the functions
$M_L(\nu)$ and $M_R(\nu)$ are to be determined from the boundary conditions.
The explicit form of the function $A(\mu)$ reads\cite{M29}
\eqb\label{A(mu)}
A(\mu) = -2\,\frac{Q_1(\mu)}{P_1(\mu)} = \frac{2}{\mu}
                                          (1-\mu\,\mbox{arctanh}\,\mu )
    = \frac{2}{\mu}\,\lambda(\mu)\,,
\eqe 
where $\lambda(\mu)$ is defined by McCormick \& 
Ku\v{s}\v{c}er \cite{MK5}.  
To apply the orthogonality relations, it is necessary to 
rewrite the solution 
in terms of the eigenfunctions used by McCormick \& 
Ku\v{s}\v{c}er\cite{MK4CZ}:
\eqb\label{phi(mu)}
\phi_{\nu}(\mu) = \frac{\nu}{2}\,{\rm P}\,\frac{1}{\nu-\mu} + \lambda(\nu)
                    \delta(\nu-\mu)\,,
\eqe
which have the property 
\eqb\label{phi(-mu)}
\phi_{-\nu}(\mu)=\phi_{\nu}(-\mu)\,.
\eqe
Equation (\ref{general}) can then be written
\eqb\label{general_neu}
  f(z,\mu) &=& a_s + 3 j\,\bigr[\mu - z(1-g_1)\bigl]
           +\int\limits_0^1\md \nu
                \tilde{M}_L(\nu)\phi_{\nu}(\mu)e^{-z/\nu} 
               +\int\limits_{0}^1\md \nu
                \tilde{M}_R(-\nu)\phi_{-\nu}(\mu)e^{(z-D)/\nu} \,,
\eqe
where we have absorbed the factor $2/\mu$ into the definition
of the functions
 $\tilde{M}_R(\mu)$ and $\tilde{M}_L(\mu)$ according to 
\eqb\label{M(mu)}
\tilde{M}_R(\mu) := \frac{2}{\mu}\, M_R(\mu)\,, \qquad\qquad 
\tilde{M}_L(\mu) := \frac{2}{\mu}\, M_L(\mu)\,.  
\eqe
The boundary conditions [Eq.~(\ref{boundary})] then become
\eqb
2\delta(\mu-\mu_0) &=&  a_s + 3 j \mu 
                   +\int\limits_0^1\md \nu
                \tilde{M}_L(\nu)\phi_{\nu}(\mu) 
               +\int\limits_{0}^1\md \nu
                \tilde{M}_R(-\nu)\phi_{-\nu}(\mu)e^{-D/\nu}
                \,,\quad \label{eq1} \\
0 &=&  a_s - 3 j \mu - 3 j D (1-g_1)
                +\int\limits_0^1\md \nu
                \tilde{M}_L(\nu)\phi_{-\nu}(\mu)e^{-D/\nu} 
               +\int\limits_{0}^1\md \nu
                \tilde{M}_R(-\nu)\phi_{\nu}(\mu)\,.\label{eq2}
\eqe
Defining
\eqb\label{B_def}
B_{\pm}(\nu) := \frac{1}{2}\,\left[\tilde{M}_L(\nu)\,
                                      \pm\,\tilde{M}_R(-\nu)\right]\,
\eqe
and adding and subtracting Eqs.~(\ref{eq1}) and (\ref{eq2}) 
leads to:
\eqb\label{together}
\delta(\mu-\mu_0) &=& \left\{\begin{array}{c} a_s \\ 3j\mu \end{array}\right\}
      \mp\frac{3}{2}jD(1-g_1)
 +\int\limits_0^1 B_{\pm}(\nu)\phi_{\nu}(\mu)\,\md\nu 
      \pm\int\limits_{0}^1 B_{\pm}(\nu)e^{-D/\nu} \phi_{-\nu}(\mu)\,\md\nu \,.
\eqe                    
In order apply the orthogonality relations, these equations
must be multiplied by a weight 
function. This function, denoted here and in McCormick \&  Ku\v{s}\v{c}er
\cite{McCoKu6} by 
$\gamma(\mu)$, is related, but not 
identical, to the $\gamma(\mu)$ defined by Mahan \cite{Mahan}, and
is given by\cite{MK15}
\eqb\label{gamma}
\gamma(\mu) = \frac{3}{2}\,\frac{\mu}{X(-\mu)}\,\,;\qquad 0\le\mu\le 1\,.
\eqe
The function $X(-\mu)$ can be written in terms of the Ambartsumian
function\cite{ambart} $\psi(\mu)$ 
or the Chandrasekhar $H$-function\cite{Chandra}. 
In the limit $c\rightarrow 1$
these relationships are\cite{Case,McCoMe}
\eqb\label{X}
X(-\mu) = \frac{\sqrt{3}}{\psi(\mu)} =  \frac{\sqrt{3}}{H(\mu)} \,.
\eqe
Tables of $X(-\mu)$, for $0\le\mu\le 1$ are given by 
Case \& Zweifel\cite{Case}; numerical
evaluation is straightforward using the representation\cite{xfunction}
\eqb
\label{explicitx}
X(-\mu) = \exp\left\lbrace {-c\over2}\int\limits_0^1 \md x 
\left(1+{c\,x^2\over 1-x^2}
\right){\ln(x+\mu)\over [1-c\,x\,{\rm arctanh}(x)]^2 + (\pi c\,x/2)^2}
\right\rbrace\,,
\eqe
where $c$ is the albedo for single scattering, equal to unity in the
case discussed here.
We now multiply Eq.~(\ref{together}) by $\gamma(\mu)$ and integrate 
over $\mu$ 
from 0 to 1. The integrals over $\mu$ can be solved using relations provided 
by  McCormick \& Ku\v{s}\v{c}er \cite{McCoKu6} (the numbers above the 
equals signs in the following refer to the relevant equation numbers):
\eqb\label{relations}
\int\limits_0^1\gamma(\mu)\,\md\mu & \stackrel{16}{=} & \gamma_0 
                                               \,\stackrel{63}{ = }\, 1 \,,\\
\int\limits_0^1\gamma(\mu)\mu\,\md\mu & \stackrel{16}{=} & \gamma_1 
      \;\stackrel{25}{ = }\; \bar{\nu}\gamma_0 \;\stackrel{63}{ = }\;\bar{\nu}
                   \;\stackrel{83}{ = }\; z_0\big|_{b=0} \;=\; 0.7104\,,\\
\int\limits_0^1\phi_{\nu}(\mu)\gamma(\mu)\,\md\mu & \stackrel{69}{=} & 0 \,,\\
\int\limits_0^1\phi_{-\nu}(\mu) \gamma(\mu)\,\md\mu & \stackrel{70}{=} & 
       \frac{3}{4}\,\frac{\nu^2}{\gamma(\nu)} \,=\, \frac{\nu}{2}\,X(-\nu)\,.
\eqe
If we denote the extrapolation distance for the Milne problem 
in the case of isotropic scattering $ z_0\big|_{b=0}=0.7104$ by
simply $z_0$, then, using the above relations, Eq.~(\ref{together})
becomes
\eqb\label{together_2}
         \frac{3}{2}\,\frac{\mu_0}{X(-\mu_0)} &=& 
                \left\{\begin{array}{c} a_s \\ 3 j z_0 \end{array}\right\} 
                                                     \mp\frac{3}{2}jD(1-g_1) 
      \pm\int\limits_{0}^1 B_{\pm}(\nu)e^{-D/\nu}\,\frac{\nu}{2}\,X(-\nu) 
                                                         \,\md\nu \,.\qquad
\eqe
The functions $B_{\pm}(\mu)$ can be calculated by multiplying 
Eq.~(\ref{together}) by $\phi_{\nu'}(\mu)\gamma(\mu)$ and integrating 
over $\mu$ from 0 to 1. Using the orthogonality relations\cite{MK6465}, 
one finds inhomogeneous Fredholm 
equations for $B_{\pm}(\mu)$ which can be solved by Neumann
iteration\cite{neumanniter}. 
In the thick slab approximation, where terms of order $e^{-D}$ 
are ignored, these Fredholm equations are trivially solved. 
Equation~(\ref{together_2}) 
for $a_s$ and $j$ is then also 
trivial and independent of $B_{\pm}(\mu)$:
\eqb\label{thick_slab}
\frac{3}{2}\,\frac{\mu_0}{X(-\mu_0)} &=& 
     \left\{\begin{array}{c} a_s \\ 3 j z_0 \end{array}\right\} 
                                          \mp\frac{3}{2}jD(1-g_1) \,.
\eqe
Once the functions $B_{\pm}(\mu)$, and hence $M_L(\mu)$ and $M_R(\mu)$  
have been found, Eq.~(\ref{thick_slab}) provides 
$a_s$ and $j$ and, therefore, the density
 $f(\mu,z)$. 
It is in principle possible to follow this procedure taking into
account higher order terms $\propto e^{-D}$. However, the 
equations become complicated in this case.

Equations~(\ref{thick_slab}) enable the transmission coefficient $T$
to be evaluated directly. 
In terms of the $X$ function we find:
\eqb\label{transmission}
T=j=\frac{\mu_0}{X(-\mu_0)}\,\frac{1}{D(1-g_1)+2z_0}\,.
\eqe
This result generalises to arbitrary $\mu_0$ ($0\le\mu_0\le 1$)
the result  of  
Mahan \cite{Mahan} [Eq.~(110)], with which it agrees for $\mu_0=1$.
In the case of isotropic scattering, $g_1=0$, 
Eq.~(\ref{transmission}) is in agreement with
the result of McCormick \& Mendelson \cite{McCoMe} [Eq.~(35)].

Finally, it should be noted that
McCormick \& Ku\v{s}\v{c}er \cite{McCoKu7} have also found 
orthogonality relations which can be used to
solve half-space transport problems with higher order anisotropy.



\begin{references}
\bibitem{MK5}Reference 7, Eq.~(5) in which, for the case of pure scattering 
              considered here, the limit $c\rightarrow 1$ must be taken.
\end{references}
\end{document}